\documentclass[aps,pra,reprint,superscriptaddress,floatfix]{revtex4-2}
\usepackage{amsmath,amssymb,bm}
\usepackage{braket}
\usepackage{graphicx}
\usepackage{hyperref}

\begin{document}

\title{Controlled Dark-State Energy Storage and Work Extraction in V-Type Quantum Batteries}

\author{Chia-Hung Chu}
\thanks{chiahungchu3030@gmail.com}
\affiliation{Institute of Atomic and Molecular Sciences, Academia Sinica, Taipei 10617, Taiwan}

\author{H. H. Jen}
\affiliation{Institute of Atomic and Molecular Sciences, Academia Sinica, Taipei 10617, Taiwan}
\affiliation{Physics Division, National Center for Theoretical Sciences, Taipei 10617, Taiwan}
\affiliation{Department of Physics, National Central University, Taoyuan City 320317, Taiwan}
\affiliation{Molecular Science and Technology Program, Taiwan International Graduate Program, Academia Sinica, Taiwan}

\date{\today}

\begin{abstract}
The performance of a quantum battery depends not only on its ability to retain stored energy against unwanted discharge but also on the amount of useful work that can be extracted from it. Here, we investigate these complementary aspects in an ensemble of V-type qutrits coupled to a lossy cavity. By applying a finite splitting pulse to the two excited states of each qutrit, cavity-bright excitations are converted into atomic dark states, whose populations remain protected upon restoring degeneracy, even while the light–matter coupling remains active. Simulations with finite-duration pulses reveal a capture window governed by the competition among energy transfer, bright-to-dark conversion, and photon leakage. We further show that if the energy is stored in the fully symmetric atomic subspace, the passive energy is bounded by twice the transition energy for number of atoms $N\geq2$. Consequently, if the stored energy scales extensively with $N$, the ratio of extractable work to stored energy approaches unity in the thermodynamic limit. Comparing different initial preparations at equal mean input energy, we find that coherent-state can yield greater finite-time ergotropy despite capturing less energy, whereas Fock-state favors greater extractable work in the stationary component. Our results demonstrate how controlled dark-state population trapping in an atom–cavity coupled system can protect stored energy while providing a conditional route toward high extractable work.
\end{abstract}

\maketitle
\raggedbottom
\section{Introduction}

The development of quantum technologies motivates controllable devices that can store energy and supply it when needed. A quantum battery realizes this concept by storing energy in controllable quantum degrees of freedom, with states and interactions manipulated during charging and extraction~\cite{Alicki2013,Campaioli2024,Ferraro2026Perspective}. Its appeal lies in using coherence, correlations, and collective control~\cite{Shukla2026,Gyhm2024,Hovhannisyan2013,Mayo2022,Francica2020,Tirone2021,Arjmandi2022} to improve energy-transfer speed and the amount of useful work recovered. Collective unitary operations can extract more work than independent operations on individual quantum batteries~\cite{Alicki2013}, highlighting the distinction between stored energy and extractable work. While stored energy quantifies the energy accumulated in a battery, ergotropy characterizes the maximum work extractable through cyclic unitary operations for a given Hamiltonian~\cite{Allahverdyan2004}. This distinction places the quantum state of the charged battery and the available control operations at the center of quantum-battery performance.

Much of the subsequent research on quantum batteries has focused on fast charging and superextensive charging power, where the total charging power grows faster than linearly with the number of batteries $N$. Theoretical studies have investigated the advantages and fundamental bounds of collective charging~\cite{Binder2015Quantacell,Campaioli2017ChargingBounds,Gyhm2022GlobalOperations,JuliaFarre2020CapacityPower},with representative proposals including common-cavity Dicke quantum batteries~\cite{Ferraro2018} and their three-level extension~\cite{Yang2024Dicke}. Superextensive charging has also been numerically demonstrated in Sachdev–Ye–Kitaev quantum batteries~\cite{Rossini2020SYKBatteries}, while a quantum charging advantage has been established in an anharmonic bosonic model~\cite{Andolina2025AnharmonicBosonic}. Further improvements in charging performance have been explored through finite-time optimal control~\cite{Mazzoncini2023Control} and reinforcement learning~\cite{Erdman2024RLDicke}. Experimentally, organic microcavity platforms have demonstrated superextensive absorption dynamics~\cite{Quach2022}, while superconducting qutrits have enabled accelerated adiabatic charging~\cite{Hu2022} and enhanced charging power through simultaneous rather than sequential driving pulses~\cite{Gemme2024QutritProtocols}. Together, these developments establish diverse routes toward rapid energy transfer and highlight the potential for quantum charging advantages over appropriately matched classical counterparts~\cite{Andolina2019QuantumClassical}.

An essential challenge in quantum batteries is to protect stored energy against unwanted discharge while preserving its availability for useful work extraction. This challenge is particularly relevant in collectively coupled systems, where strong light–matter interactions facilitate rapid energy transfer but can also enhance radiative losses through superradiant emission~\cite{Dicke1954,Tibben2025}. Dissipation in the charger further influences the balance between transferred energy and ergotropy~\cite{Farina2019}. Dark states offer a promising route to suppressing these losses~\cite{Song2024} by trapping excitations in nonradiative subspaces.~\cite{Lidar1998}. Such strategies have been explored through stabilized three-level charging~\cite{Santos2019}, symmetry-protected storage in atomic networks~\cite{Liu2019}, reservoir-engineered collective dark states~\cite{Quach2020} and superconducting-qutrit experiments~\cite{Hu2022}. Alternative approaches include transferring excitations into long-lived molecular states~\cite{Tibben2025}, exploiting Floquet bound states~\cite{BaiAn2020Floquet} and topological reservoir~\cite{Lu2025}, and employing measurement-based control~\cite{Gherardini2020Sequential,Mitchison2021Feedback,Cenedese2026} or localization~\cite{Rossini2019LocalizedBatteries}. However, dark-state protection introduces a fundamental trade-off: the symmetry that suppresses radiative decay can also restrict access to the stored excitations. This raises a central question: Can dark-state populations be dynamically controlled without compromising their protection or the extractable work? Atomic permutation symmetry provides a particularly useful framework, as it confines the dynamics to a fully symmetric subspace whose dimension grows only polynomially with system size. When the stored energy scales extensively, this restricted structure can suppress correlation-induced work locking, allowing the ratio of extractable work to stored energy to approach unity in the thermodynamic limit\cite{Andolina2019}. Conversely, breaking permutation symmetry can populate additional symmetry sectors and potentially reduce ergotropy. These considerations motivate a controlled dark-state storage protocol that enables access to protected excitations while preserving atomic permutation symmetry and maintaining substantial extractable work.

Here we investigate such a protocol for an ensemble of $N$ identical V-type qutrits coupled to a single cavity. The cavity initially excites collective bright states, while a finite splitting pulse applied to the two excited states of each qutrit transfers population into atomic dark states. Restoring the excited-state degeneracy protects the captured dark population even while the cavity coupling remains active. Crucially, the pulse temporarily breaks the exchange symmetry between the two excited states of each qutrit but acts uniformly across the ensemble, preserving atomic permutation symmetry throughout the protocol. We connect this protection mechanism to a bound on passive energy, establishing a condition under which the ratio of extractable work to stored energy approaches unity in the thermodynamic limit. Finite-pulse simulations further reveal the competition among charging, bright-to-dark conversion, and photon leakage, while comparing Fock-state and coherent-state preparations in terms of retained energy and atomic ergotropy. Section~\ref{sec:model} introduces the model, control protocol, and work measure. Section~\ref{sec:results} develops the conservation law and extractability bound, then analyzes finite-time capture and the stationary component. Section~\ref{sec:discussion} discusses the implications, operating compromises, and future extensions.

\section{Model and storage protocol}
\label{sec:model}
\subsection{Cavity-coupled V-type ensemble}

We consider an ensemble of $N$ identical V-type qutrits coupled to a single-mode cavity. The qutrits constitute the quantum battery, while the cavity mediates energy transfer and provides a photon-loss channel. Equal and in-phase coupling to the two excited states enables excitation through the cavity-bright sector, whereas the orthogonal atomic dark states provide a protected subspace for energy storage. A temporary splitting between the two excited states controls bright-to-dark population transfer while the cavity coupling remains active.

Each qutrit has a ground state $\ket{g_n}$ and two excited states $\ket{e_{n,1}}$, $\ket{e_{n,2}}$, degenerate at energy $\omega$ when the pulse is off. With $\hbar=1$ and the atomic ground energy set to zero, the Hamiltonian can be expressed as:
\begin{align}
 H(t)&=H_c+H_a+H_{\rm int}+H_p(t),\nonumber\\
 H_c&=\omega \hat a^\dagger \hat a,\qquad
 H_a=\omega(\hat n_{e_1}+\hat n_{e_2}),\label{eq:bare_hamiltonian}\\
 H_{\rm int}&=\frac{g_0}{\sqrt{2N}}(\hat a+\hat a^\dagger)
 \sum_{n=1}^N\sum_{j=1}^2
 \bigl(\ket{g_n}\bra{e_{n,j}}+\mathrm{H.c.}\bigr),\label{eq:interaction}\\
 H_p(t)&=\frac{\delta(t)}{2}(\hat n_{e_1}-\hat n_{e_2}).
 \label{eq:pulse}
\end{align}
Here $\omega$, $g_0$, and $\hat n_{e_j}=\sum_n\ket{e_{n,j}}\bra{e_{n,j}}$ are the resonant frequency, interaction strength, and number operator of excited states. $\hat a$ ($\hat a^\dagger$) are the photon annihilation (creation) operators. The four terms in the Hamiltonian correspond to the cavity, atoms, interactions, and the controlled pulse, respectively. There are no direct interactions between qutrits and coupling is present throughout charging and storage; only the splitting $\delta(t)$ is varied.

Define the bright and dark atomic states as follows:
\begin{equation}
 \ket{B_n}=\frac{\ket{e_{n,1}}+\ket{e_{n,2}}}{\sqrt2},\qquad
 \ket{D_n}=\frac{\ket{e_{n,1}}-\ket{e_{n,2}}}{\sqrt2}.
 \label{eq:bright_dark}
\end{equation}
Writing $\hat N_B=\sum_n\ket{B_n}\bra{B_n}$ and $\hat N_D=\sum_n\ket{D_n}\bra{D_n}$ gives $H_a=\omega(\hat N_B+\hat N_D)$ and
\begin{align}
 H_{\rm int}&=\frac{g_0}{\sqrt N}(\hat a+\hat a^\dagger)
 \sum_n\bigl(\ket{g_n}\bra{B_n}+\mathrm{H.c.}\bigr),\label{eq:bright_coupling}\\
 H_p(t)&=\frac{\delta(t)}2\sum_n
 \bigl(\ket{B_n}\bra{D_n}+\mathrm{H.c.}\bigr).
 \label{eq:bright_dark_pulse}
\end{align}
In the absence of the control pulse, $H_c+H_a+H_{\rm int}$ couples the cavity only to the bright transition: the dark state is decoupled from cavity exchange, and its population is conserved. The splitting pulse induces bright–dark mixing, enabling population transfer into the protected storage channel. We keep the collective coupling strength $g_0$ fixed as $N$ varies, such that the single-qutrit bright-transition coupling scales as $g_0/\sqrt{N}$.

We describe the time evolution of the total atom-cavity density operator $\rho(t)$ using a Lindblad master equation under the Markovian assumption that the reservoir has no memory on the system's evolution timescale~\cite{Lindblad1976,CohenTannoudji1998}. The dissipative process considered here is photon leakage into the environment, represented by the local cavity jump operator $L=\sqrt\kappa\,\hat a$:
\begin{equation}
 \dot\rho=\mathcal L(t)\rho
 =-i[H(t),\rho]+\kappa\left(\hat a \rho \hat a^\dagger
 -\tfrac12\{\hat a^\dagger \hat a,\rho\}\right).
 \label{eq:lindblad}
\end{equation}
We retain the counter-rotating terms in $H_{\mathrm{int}}$ to describe coupling strengths up to $g_0 \simeq 0.3\omega$ beyond the rotating-wave approximation. These terms couple the bare vacuum–ground state to configurations containing photons and bright excitations. Consequently, within the local photon-loss model considered here, the coupled bright–cavity sector does not necessarily relax to the bare vacuum–ground product state at long times~\cite{Beaudoin2011}.

\subsection{State initializations, control protocol, and extractable work}
\label{sec:protocol}

Initially, the atoms occupy the many-body ground state $\ket G=\ket g^{\otimes N}$, and the cavity is either in the Fock state $\ket N$ or the coherent state $\ket{\alpha=\sqrt N}$. The Fock and coherent cavity states have the same mean photon number $N$, from which we can directly compare their storage performance at equal mean input energy.

When a splitting pulse is applied to the two excited states, it induces coherent mixing between the bright and dark states. This dynamics can be understood from an isolated qutrit: in the interaction picture with respect to $H_a$, the splitting drives a rotation between $|B\rangle$ and $|D\rangle$. For a square pulse of amplitude $\delta$ and duration $\tau_p$ on the bright state, a single qutrit evolves as:
\begin{equation}
 U_p(\tau_p,0)\ket B
 =\cos\!\left(\frac{\delta\tau_p}{2}\right)\ket B
 -i\sin\!\left(\frac{\delta\tau_p}{2}\right)\ket D.
 \label{eq:ideal_transfer}
\end{equation}
where $U_p(\tau_p,0) = e^{-iH_p\tau_p}$ is the time evolution corresponding to the pulse Hamiltonian $H_p$ (Eq.~\eqref{eq:pulse}). A pulse area of $\delta\tau_p=\pi$ achieves complete population transfer from the bright state to the dark state, up to an overall phase. Accordingly, we employ a square splitting pulse with $\delta>0$ and duration $\tau_p=\pi/\delta$.

The isolated-qutrit rotation provides a useful guide for controlling the coupled system. When the pulse-driven rotation is much faster than the light–matter energy exchange and cavity loss, their effects during the pulse are negligible. Bright-state population can therefore be accumulated through cavity-mediated charging and subsequently transferred into dark states by a rapid splitting pulse. For finite pulse durations, however, energy exchange and photon leakage compete with bright-to-dark conversion, and both processes are fully accounted for in our numerical simulations.

To maximize the bright-state population available for conversion, we adopt a bright-peak timing criterion. We define $t_{B,\rm peak}$ as the time at which the pulse-free bright-state population $\braket{\hat N_B}^{(0)}_t$ reaches its sampled maximum within the charging interval. The splitting pulse is then centered at $t_{B,\rm peak}$ whenever possible, with its onset and termination times given by
\begin{equation}
 t_{\rm on}=\max(0,t_{B,\rm peak}-\tau_p/2),\qquad
 t_{\rm off}=t_{\rm on}+\tau_p.
 \label{eq:timing}
\end{equation}
The pulse timing is determined independently for each initial state and parameter set. When $t_{B,\rm peak}<\tau_p/2$, the pulse starts at $t=0$, allowing charging and bright-to-dark conversion to occur simultaneously. This prescription provides a physically motivated criterion for capturing bright-state excitations. Unless otherwise specified, we apply the bright-peak timing criterion throughout the remaining simulations.

For $\rho_a=\operatorname{Tr}_c(\rho)$, we evaluate the bare atomic energy $E_a=\operatorname{Tr}(H_a\rho_a)$ and ergotropy~\cite{Allahverdyan2004}
\begin{align}
 \mathcal E(\rho_a)&=E_a-E_{\rm pass},\nonumber\\
 E_{\rm pass}&=\min_{U\in\mathrm U(\mathcal H_a)}
 \operatorname{Tr}(H_aU\rho_a U^\dagger).
 \label{eq:ergotropy}
\end{align}
Here $U$ acts on the atomic tensor-product space $\mathcal H_a = (\mathbb C^3)^{\otimes N}$. The passive energy pairs the decreasing eigenvalues of $\rho_a$ with the increasing energies of $H_a$ in this space. Ergotropy thus separates the stored atomic energy from the part that cannot be recovered by a cyclic unitary operation. We also use dark and bright populations $\braket{\hat{N}_D}$, $\braket{\hat{N}_B}$, and the atomic purity $\operatorname{Tr}(\rho_a^2)$ to characterize the stored state.

\begin{figure}[b]
 \centering
 \includegraphics[width=\linewidth]{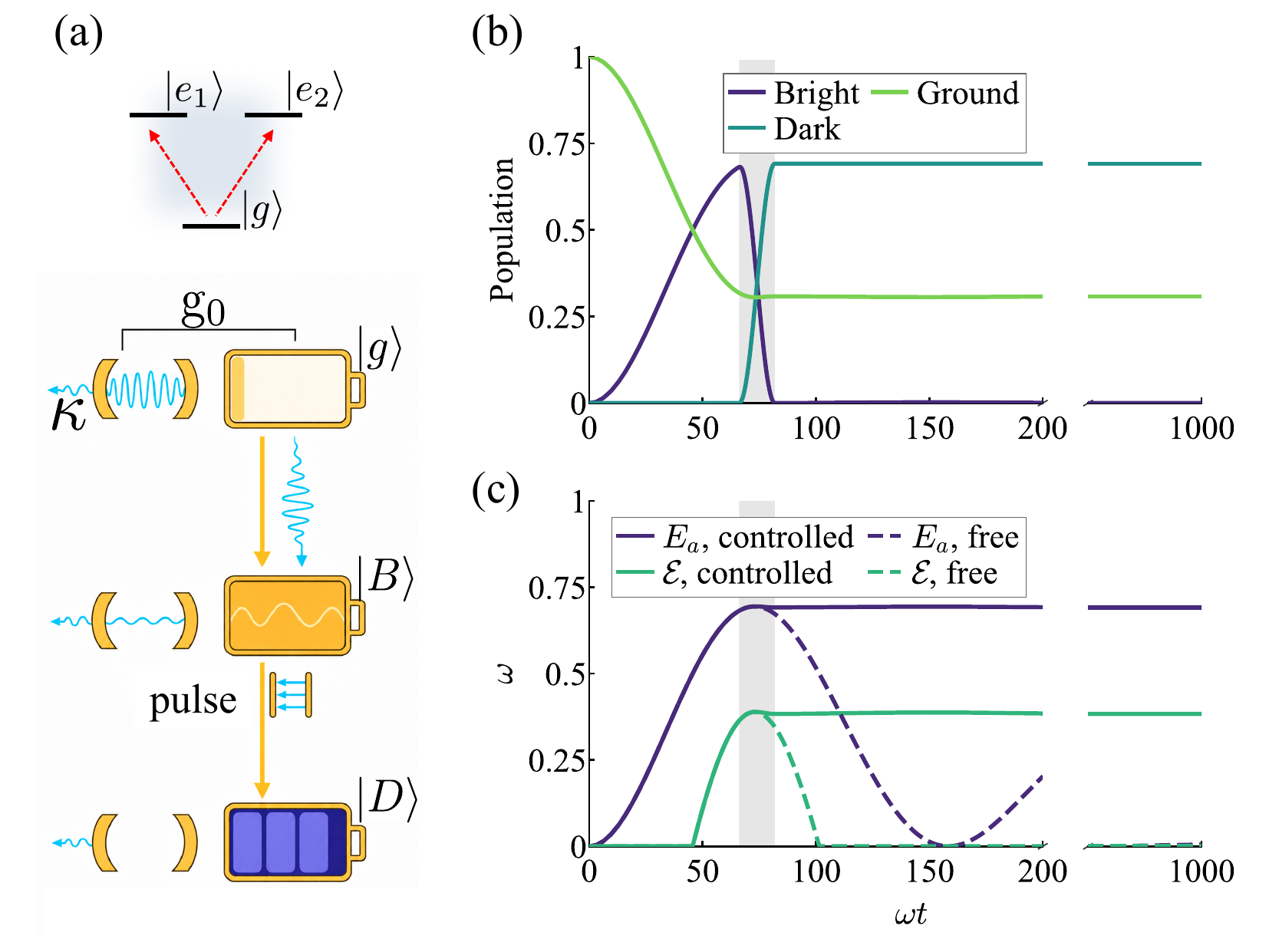}
 \caption{Bright-to-dark storage. (a) V-type levels and the cavity-charging and pulse-capture sequence. (b) Bright, dark, and ground state populations for a single qutrit. (c) Atomic energy and ergotropy with the pulse (solid) and without it (dashed). The cavity initially prepared in a single photon Fock state $\ket{1}$, while the qutrit is in the ground state $\ket{g}$, with $g_0=0.02\omega$ and $\kappa=0.01\omega$. The gray interval marks a pulse with $\delta=0.2\omega$, duration $\pi/\delta$, and centered at $\omega t_c=74$, where $t_c=(t_{\rm on}+t_{\rm off})/2$. Panels (b,c) show the time evolution in a range of $0\leq\omega t\leq200$ and $950\leq\omega t\leq1000$, separated by an axis break.}
 \label{fig:mechanism}
\end{figure}
\section{Results}
\label{sec:results}
\subsection{Bright-to-dark transfer and protected population}
\label{sec:mechanism}

Our protocol converts bright-state excitation into protected dark-state population, enabling energy storage beyond the transient charging process. As illustrated schematically in Fig.~\ref{fig:mechanism}(a), cavity-mediated charging first populates the bright state, followed by a splitting pulse that transfers excitation into the dark state. In the single-qutrit example shown in Fig.~\ref{fig:mechanism}(b), the bright-state population builds up before the pulse and is subsequently converted into dark-state population that remains constant after the pulse ends. The energetic consequence is demonstrated in Fig.~\ref{fig:mechanism}(c), where the controlled qutrit retains finite energy and ergotropy at late times, in stark contrast to the pulse-free dynamics, which exhibits negligible energy and vanishing ergotropy. Notably, the difference between the retained energy and ergotropy highlights the distinction between protecting excitation and preserving extractable work.

The protection of the dark-state population follows directly from the pulse-off Liouvillian. With $H_{\rm off}=H_c+H_a+H_{\rm int}$,
\begin{equation}
 [H_{\rm off},\hat N_D]=0,\qquad [a,\hat N_D]=0,
 \label{eq:dark_commutators}
\end{equation}
which implies $\mathcal L_{\rm off}^{\dagger}\hat N_D=0$ and hence
\begin{equation}
 \braket{\hat{N}_D}_t=\braket{\hat{N}_D}_{t_{\rm{off}}}.
 \label{eq:dark_conservation}
\end{equation}
Furthermore, every projector onto a fixed-$N_D$ sector is conserved as well. This conservation guarantees the protection of the energy contribution $\omega\braket{\hat N_D}$, but does not necessarily preserve the full atomic density matrix or its ergotropy. A natural question then arises: under what conditions does this protected energy also retain extractable work?

\subsection{Work extractability of the dark-state population}
\label{sec:rank_bound}

Permutation symmetry provides a direct connection between protected dark-state population and many-body extractability. Since the initial atomic state is fully symmetric, all Hamiltonian terms preserve the symmetric subspace, and the cavity jump operator acts trivially on the atomic degrees of freedom, the Lindblad evolution remains confined to the fully symmetric qutrit subspace. Consequently, the reduced atomic density matrix satisfies the rank bound
\begin{equation}
 \operatorname{rank}\rho_a\leq d_{\rm sym}=\binom{N+2}{2}.
 \label{eq:rank}
\end{equation}
This symmetry-imposed constraint provides a basis for establishing bounds on extractable work, even when the atomic battery is in a mixed state.

To establish a lower bound on ergotropy, we consider the passive rearrangement in Eq.~\eqref{eq:ergotropy}, which involves the full atomic energy spectrum. The many-body energy level $q\omega$ has a degeneracy
\begin{equation}
 d_q=\binom Nq 2^q,\qquad q=0,\ldots,N.
\end{equation}
For $N\geq2$, the ground state and the first two excitation manifolds together accommodate
\begin{equation}
 d_0+d_1+d_2=1+2N+4\binom N2
 =1+2N^2\geq d_{\rm sym}.
 \label{eq:level_count}
\end{equation}
Combined with the symmetry-imposed rank bound in Eq.~\eqref{eq:rank}, this ensures that all nonzero eigenvalues of $\rho_a$ can be rearranged into energy levels no higher than $2\omega$ in the corresponding passive state. Consequently, the passive energy is bounded by $E_{\mathrm{pass}}\leq 2\omega$, yielding
\begin{equation}
 E_{\rm pass}\leq2\omega,\qquad
 \mathcal E\geq\max\{0,E_a-2\omega\}.
 \label{eq:passive_bound}
\end{equation}
This result extends the rank-based reasoning for extractable work developed by Andolina \textit{et al.}~\cite{Andolina2019} to the present symmetric open quantum system, demonstrating how permutation symmetry constrains the passive energy and guarantees a finite amount of extractable work whenever $E_a>2\omega$.

Because $E_a\geq \omega\braket{\hat{N}_D}$, the conserved dark-state population establishes a time-independent lower bound on ergotropy after the pulse
\begin{equation}
 \mathcal E(t)\geq\max\{0,(\braket{\hat{N}_D}_{t_{\rm off}}-2)\omega\}.
 \label{eq:dark_work_bound}
\end{equation}
For ${\braket{\hat{N}_D}}>0$, the fraction of extractable energy further satisfies
\begin{equation}
 1\geq\frac{\mathcal E}{E_a}
 \geq1-\frac{2\omega}{E_a}
 \geq1-\frac{2}{\braket{\hat{N}_D}}.
 \label{eq:extractability_bound}
\end{equation}
Remarkably, when the retained energy scales extensively with the battery size, $E_a\propto N\omega$, the passive energy remains bounded by $2\omega$, independent of $N$. Consequently, the extractable fraction approaches unity in the thermodynamic limit, demonstrating that dark-state population protection, together with permutation symmetry, enables asymptotically complete extraction of the stored energy.

\begin{figure}[htbp]
 \centering
 \includegraphics[width=\linewidth]{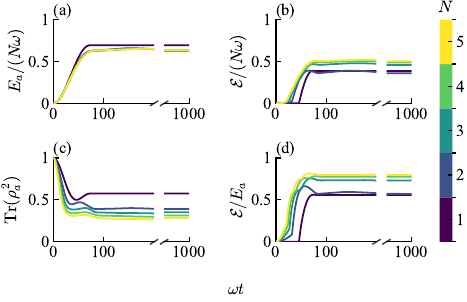}
 \caption{Stored energy and extractability as the battery size increases. Time evolution of (a) atomic energy per qutrit, (b) ergotropy per qutrit, (c) atomic purity, and (d) extractable fraction for $N$-qutrits battery with input Fock states $\ket{N}$, $N=1,\ldots,5$ indicated by the shared colorbar. Parameters are $g_0=0.02\omega$, $\kappa=0.01\omega$, and $\delta=0.2\omega$, with bright-peak timing for each $N$. The windows for time evolution are chosen as $0\leq\omega t\leq200$ and $950\leq\omega t\leq1000$, for short- and long-time limits.}
 \label{fig:extractability}
\end{figure}

The four observables in Fig.~\ref{fig:extractability} distinguish energy retention from the quality of the stored state. For $N=2,\ldots,5$, the late-time energy per qutrit changes little, while both the ergotropy per qutrit and the extractable fraction increase, in Figs.~\ref{fig:extractability}(b) and ~\ref{fig:extractability}(d), respectively. This improvement occurs even as the atomic purity decreases as shown in Fig.~\ref{fig:extractability}(c). The mixedness of the atomic state alone does not determine how much energy is locked in its passive component. The finite-size trajectories further illustrate this distinction between stored energy and extractable work, consistent with the symmetry-based bound. Importantly, the conclusion of asymptotically complete extractability remains valid in the large-$N$ limit, provided that the retained energy scales extensively with the battery size.

\subsection{Finite-pulse capture and its control-dependent operating range}
\label{sec:capture}

\begin{figure}[htbp]
 \centering
 \includegraphics[width=\linewidth]{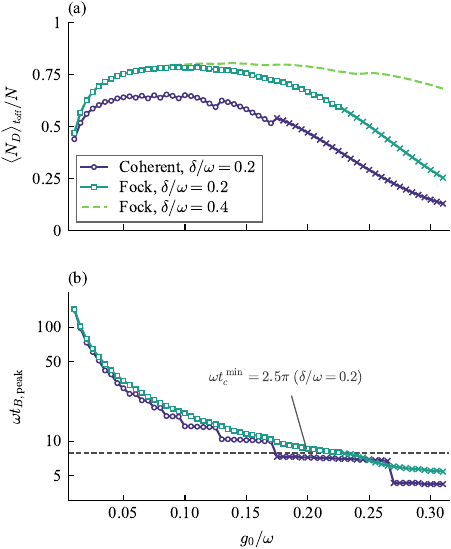}
 \caption{Capture and the charging timescale. (a) Captured dark population for input Fock and coherent state at $\delta=0.2\omega$; the dashed curve for input Fock state at $\delta=0.4\omega$ is a reference for faster conversion. All pulses have area $\pi$. (b) Free evolution bright-peak times for the two initial states, shown on a logarithmic time axis. The dashed horizontal line is the minimum pulse-center time for $\delta=0.2\omega$. Crosses in both panels replace the corresponding circle or square when the peak precedes that line and $t_{\rm on}=0$. Parameters are $N=3$ and $\kappa=0.01\omega$, with 61 samples of $g_0$ scanned from $[0.01,0.31]\omega$.}
 \label{fig:capture}
\end{figure}

While dark-state protection ensures the retention of captured excitation, the efficiency of bright-to-dark transfer depends critically on the finite duration of the control pulse. In particular, the pulse must act sufficiently rapidly to transfer excitation before cavity-mediated exchange and photon leakage substantially modify the bright-state population available for capture. Figure ~\ref{fig:capture}(a) illustrates the resulting operating range as the collective coupling strength is varied. At weak coupling, slow charging allows photons to escape before a substantial bright-state population develops. As the coupling increases, charging becomes faster, but cavity-driven dynamics increasingly compete with the finite-duration bright-to-dark rotation. Consequently, the capture efficiency is maximized at an intermediate coupling strength, near $g_0=0.1\omega$ for the pulse parameters considered here.

The peak times shown in Fig.~\ref{fig:capture}(b) reveal an additional limitation at stronger coupling. For $\delta=0.2\omega$, the bright-state population in the pulse-free evolution eventually reaches its maximum before the earliest possible pulse center, $\omega\tau_p/2=2.5\pi$. The crosses mark the parameter points where the timing prescription in Eq.~\eqref{eq:timing} consequently sets $t_{\rm on}=0$, causing charging and bright-to-dark conversion to occur simultaneously.  The resulting reduction in capture efficiency therefore reflects both the competition between cavity dynamics and finite-duration control, and the constraint imposed by the pulse-timing boundary. Importantly, this behavior characterizes the operating range of the chosen control protocol rather than an intrinsic upper limit on the useful coupling strength.

The dashed reference curve in Fig.~\ref{fig:capture}(a) further supports this timescale interpretation. Increasing the pulse amplitude while maintaining a fixed pulse area shortens the conversion time, shifts the optimal capture toward stronger coupling, and improves the capture efficiency in regimes where the slower pulse becomes less effective. 

Experimentally, the required splitting may be implemented through a differential Stark or Zeeman shift. We consider the perturbative regime $\delta\ll\omega$, which maintains the separation between the ground and excited manifolds. Arbitrarily increasing $\delta$ to approach an instantaneous $\pi$ rotation would violate this assumption. Thus, particularly in the strong-coupling regime, bright-to-dark conversion remains a finite-duration process that competes with ongoing cavity-mediated charging. Having identified the operating range for efficient population capture, we next examine how cavity loss influences the retention of stored energy and extractable work, taking $g_0=0.1\omega$ as a representative coupling strength.

\subsection{Loss-dependent capture and finite-time extractable work}
\label{sec:loss}

\begin{figure}[htbp]
 \centering
 \includegraphics[width=\linewidth]{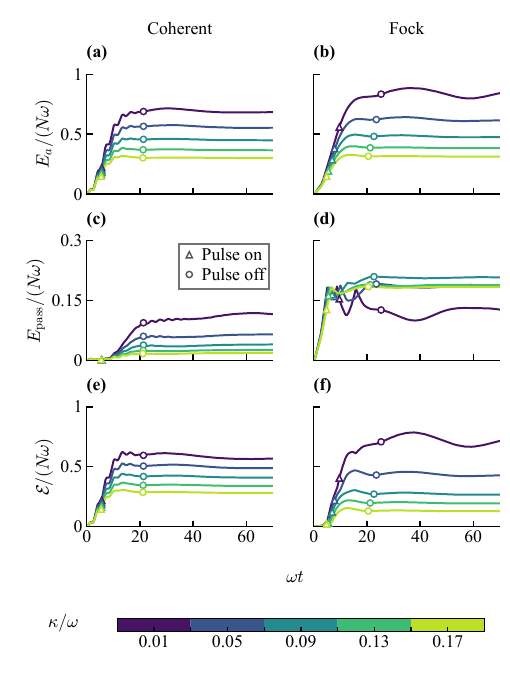}
 \caption{Finite-time energy and extractable work under cavity loss. The left and right columns correspond to coherent-state and Fock-state inputs, respectively. The rows show (a,b) atomic energy $E_a$, (c,d) passive energy $E_{\mathrm{pass}}=E_a-\mathcal{E}$, and (e,f) ergotropy $\mathcal{E}$, all normalized by $N\omega$. The shared colorbar indicates the cavity loss rate $\kappa/\omega.$ Triangles and open circles mark the pulse-on and pulse-off times, respectively, along each trajectory. Note the reduced vertical scale for the passive-energy panels. Parameters are $N=3$, $g_0=0.1\omega$, and $\delta=0.2\omega$, with the same nominal input energy of $3\omega$ for both initial preparations. The pulse timing is determined separately for each case using the bright-peak criterion. Time is measured from the common initial preparation.}
 \label{fig:loss}
\end{figure}

Cavity loss influences the protocol differently before and after bright-to-dark transfer. Before and during the control pulse, photon leakage depletes the excitation available for capture. After the pulse, cavity loss continues to affect the remaining coupled component, while the captured dark-state population remains protected. Figure~\ref{fig:loss} illustrates these distinct effects by comparing the atomic energy, passive energy, and ergotropy for coherent-state and Fock-state inputs with the same nominal energy. Both preparations employ identical pulse amplitudes and areas, while the pulse timing is optimized separately for each initial state according to the bright-peak criterion.

A key observation from Fig.~\ref{fig:loss} is that greater stored energy does not necessarily translate into more extractable work at finite times. Comparing the atomic energy and ergotropy in Fig.~\ref{fig:loss} reveals that, at the end of the plotted time window, the Fock-state input retains more atomic energy at every sampled cavity loss rate, whereas the coherent-state input yields greater ergotropy at the four larger loss rates. The pulse markers further indicate that both preparations are compared at the same physical time, despite their distinct pulse boundaries determined by the bright-peak criterion. This apparent reversal can be understood from the passive-energy contributions in Figs.~\ref{fig:loss}(c) and ~\ref{fig:loss}(d): the smaller passive energy associated with coherent-state input can compensate for its lower stored energy, resulting in greater extractable work. Thus, the initial cavity-field preparation influences not only the amount of energy stored in the atomic battery but also its partition into passive energy and ergotropy. This finite-time advantage of coherent-state input is consistent with the reduced correlation-induced energy locking reported in previous works for closed-system charging~\cite{Andolina2019,Yang2024Dicke}, where the comparison was performed at the time of maximum average power. However, a transient advantage in ergotropy does not necessarily translate into better long-term storage under Lindblad evolution. An important question then arises: how much of this extractable work survives in the stationary component after the bright-to-dark transfer?

\subsection{Preparation ranking in the stationary component}
\label{sec:stationary}

To examine the extractable work retained in the long-time limit, we consider the stationary component of the post-pulse Liouvillian $\mathcal L_{\rm off}$, which becomes time independent after the control pulse. Its projector $\mathsf P_0$ onto the zero-eigenvalue subspace is constructed from the biorthogonal left and right eigenoperators of $\mathcal L_{\rm off}$~\cite{Albert2014}. We define
\begin{align}
 \rho_0&=\mathsf P_0\rho(t_{\rm off}),&
 \rho_{a,0}&=\operatorname{Tr}_c(\rho_0),\nonumber\\
 \mathcal E_0&=\mathcal E(\rho_{a,0}),&
 E_{a,0}&=\operatorname{Tr}(H_a\rho_{a,0}).
 \label{eq:zero_projection}
\end{align}
Here, the subscript $0$ denotes the stationary component, which represents the full long-time limit provided that all other populated Liouvillian modes decay. In particular, nonzero purely imaginary eigenvalues may sustain persistent oscillations beyond the zero-mode projection.

\begin{figure}[t]
 \centering
 \includegraphics[width=\linewidth]{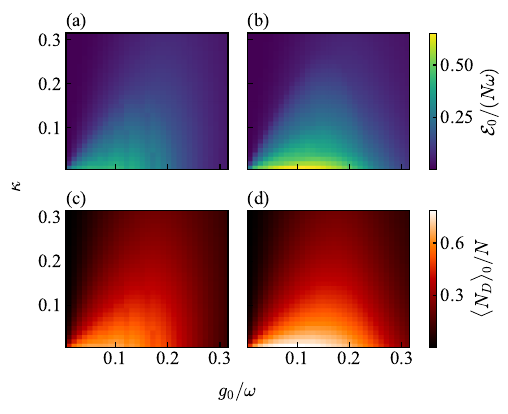}
 \caption{Stationary extractable work and protected dark-state population. The left and right columns correspond to coherent-state and Fock-state inputs, respectively. Panels (a,b) show the zero-mode-projected atomic ergotropy $\mathcal{E}_0/(N\omega)$, while panels (c,d) display the normalized dark-state population $\braket{\hat N_D}_0/N$, which equals its value at pulse-off due to conservation. Each row shares a common color scale across both preparations. Parameters are $N=3$ and $\delta=0.2\omega$, with pulse timing determined by the bright-peak criterion on a $31\times31$ parameter grid, where $g_0/\omega$,$\kappa/\omega=0.01,0.02$,$\ldots,0.31$. All 961 parameter points are evaluated for each preparation, and cell colors represent the calculated values without interpolation. The vertical axis denotes $\kappa/\omega$, with $\omega=1$ in the numerical data.}
 \label{fig:stationary}
\end{figure}

Figure~\ref{fig:stationary} compares this stationary work with the protected dark-state population. At the representative coupling $g_0=0.1\omega$, Fock-state preparation yields both greater captured dark-state population and larger stationary ergotropy $\mathcal E_0$ at all five loss rates used in Fig.~\ref{fig:loss}. Since the two figures employ identical pulse windows at these parameter points, this comparison reveals that the finite-time ergotropy advantage of coherent-state preparation observed in Fig.~\ref{fig:loss} does not persist in the stationary component.

The two rows of Fig.~\ref{fig:stationary} further illustrate the relationship between population capture and stationary extractable work across the operating range. Both quantities are enhanced at weak cavity loss and intermediate coupling strengths, where excitation can be efficiently charged and transferred into the dark state before photon leakage becomes dominant. Comparing Fig.~\ref{fig:stationary}(a) and ~\ref{fig:stationary}(b), Fock-state preparation generally yields greater stationary ergotropy in the region where substantial extractable work is retained. The corresponding dark-state populations in Figs.~\ref{fig:stationary}(c) and ~\ref{fig:stationary}(d) exhibit a similar advantage in capture efficiency. These results demonstrate that the initial field preparation yielding greater finite-time extractable work does not necessarily maximize the work retained in the stationary component. This conclusion applies to the sampled $N=3$ parameter grid and the specified finite-pulse control protocol.
\section{Discussion and conclusions}
\label{sec:discussion}

We have proposed a quantum-battery charging protocol that separates energy charging from protected storage through coherent control of the battery’s internal states. Cavity-mediated charging first populates the bright state, followed by a finite-duration splitting pulse that transfers excitation into the dark state. Restoring the excited-state degeneracy after the pulse protects the captured dark-state population even while the light–matter coupling remains active. This fixed-coupling setting is motivated by an experimentally demonstrated multilayer molecular battery~\cite{Hymas2026}, in which an integrated microcavity combines the energy-storage medium and charge-transport layers within a single device. Such an architecture motivates external control of internal energy levels to achieve protected storage without reconfiguring the cavity–battery structure.

Previous studies have explored multilevel structures and protected states through different mechanisms. Adiabatic three-level protocols follow driven ladder dark states to achieve stable charging~\cite{Santos2019,Hu2022}, while three-level Dicke batteries investigate cavity charging in a cascade configuration~\cite{Yang2024Dicke}. In contrast, our V-type configuration enables a finite internal rotation of excitation. Other approaches exploit network symmetry and selective environmental coupling to protect stored excitation, with symmetry-breaking perturbations enabling its release~\cite{Liu2019}, or prepare collective charger–battery dark states through a shared reservoir~\cite{Quach2020}. Molecular triplet states have also been employed to separate rapid absorption from energy retention in microcavity experiments. Our protocol instead uses a uniform splitting pulse to achieve controlled internal capture while preserving ensemble permutation symmetry and maintaining an unchanged cavity coupling.

A central contribution of this work is the connection between controlled population protection and many-body extractable work. Two symmetries play complementary roles: excited-state exchange symmetry defines the cavity-dark state, while ensemble permutation symmetry restricts the rank of the atomic density matrix. The collective Hamiltonian and cavity-only dissipation preserve the fully symmetric atomic subspace even when the battery becomes mixed. Extending the rank-based extractability argument ~\cite{Andolina2019} to the present open qutrit model yields $E_{\rm pass}\leq2\omega$ for $N\geq2$. When the retained energy scales extensively with the battery size, this size-independent passive-energy bound implies $\mathcal E/E_a\to1$ as $N\to\infty$. The conserved dark-state population further establishes a time-independent lower bound on atomic ergotropy after the pulse. Together, internal-state capture and restricted many-body support provide a criterion for retaining useful work while the battery remains coupled to its charger.

The efficiency of the protocol is constrained by competition among finite-duration control, cavity-mediated exchange, and photon leakage. This competition motivates pulse shaping and timing optimization for a specified storage objective. The initial cavity-field preparation introduces an additional tradeoff: coherent-state input can yield greater finite-time ergotropy through a smaller passive-energy contribution, whereas Fock-state input favors greater zero-mode-projected ergotropy in the parameter regimes studied. The projected state represents the full long-time state only when all other populated Liouvillian modes decay ~\cite{Albert2014}. The protection mechanism relies on equal light–matter couplings, restored excited-state degeneracy, and cavity-only dissipation. Investigating unequal couplings and atomic noise~\cite{Canzio2025LocalNoise} clarify the lifetime of the captured population and the robustness of the symmetry-based ergotropy bound. Additional dissipative channels, including dephasing, have also been considered in quantum-battery models~\cite{Shen2026Nonlinear,Shastri2025}, although their treatment requires a bath description consistent with the relevant light–matter coupling regime~\cite{Beaudoin2011}. Finally, the global atomic ergotropy benchmark assumes ideal unitary work extraction~\cite{Alicki2013,Andolina2019}. Incorporating an explicit load stage, as demonstrated experimentally in NMR quantum batteries~\cite{JoshiMahesh2022NMR}, would connect the retained work to delivered energy~\cite{Niu2024}. Accounting for control and extraction costs, including changes in interaction energy~\cite{Ferraro2026Perspective}, would enable a comprehensive assessment of the protocol over a complete operating cycle.
\begin{acknowledgments}

We acknowledge support from the National Science and
Technology Council (NSTC), Taiwan, under the Grants No.
NSTC-115-2112-M-001035-MY3 and No. NSTC-115-2119-
M-001-009, and from Academia Sinica under Grant AS-
CDA-113-M04. We are also grateful for support from TG 1.2
of NCTS at NTU.

\end{acknowledgments}

\section*{Data Availability}
The data that support the findings of this article are not pub-
licly available. The data are available from the authors upon
reasonable request.
\bibliography{reference}
\end{document}